# Vacuum permittivity and gravitational refractive index, revisited

Ion Simaciu[*]

Retired lecturer, Petroleum-Gas University of Ploiești, Ploiești 100680, Romania

E-mail: [*] isimaciu@yahoo.com, ion.simaciu@gmail.com

Abstract:
The present paper reanalyzes the problem of the refractive properties of the physical vacuum and their modification under the action of the gravitational field and the electromagnetic field. This problem was studied in our previous works and in the subsequent works of the researchers: Leuchs, Urban, Mainland and their collaborators. By modeling the physical vacuum as a particle-antiparticle system, we can deduce with a certain approximation, in a semiclassical theory, the properties of the free vacuum and the vacuum modified by the interaction with a gravitational field and an electromagnetic field. More precise calculation of permittivities of free vacuum and near a particle can lead to a non-point model of the particle. This modeling can follow both the quantum and the general relativistic path as well as the phenomenological path, the results complementing each other.

## 1. Introduction

In the paper "Radiation - medium - curvature interaction" [1] we tried to model physico-mathematically the phenomenon of dispersion and absorption determined by the gravitational interaction between real particles in our universe, i.e. the electromagnetic world (EMW) and pairs of virtual particles in the physical vacuum (PV). The present paper synthesizes and corrects the study carried out in previous papers [1 - 4] and the doctoral thesis [5] on the emergence of relativistic and quantum properties from the interaction between real particles and the vacuum modeled as a superconducting and superfluidic physical medium [1, 6].

The paper [1] is a development of the papers of Wilson [7], Dicke [8] and Kar [9] on modeling the gravitational interaction as an electromagnetic phenomenon. According to this hypothesis, to explain the gravitational interaction of electrically charged systems and systems composed of these particles it is sufficient to accept that it modifies the refractive index of the physical vacuum, considered as a dielectric and paramagnetic, through the phenomena of polarization and magnetization. This modification of the refractive index of the vacuum determines the modification of the propagation speed of electromagnetic waves (emw), $c = c_0 \varepsilon_{rg}^{-1}$, $c_0 = 1/\sqrt{\varepsilon_0 \mu_0}$ of the energy of electromagnetic systems, $E = E_0 \varepsilon_{rg}^{-1/2}$ and of the other parameters: the spatial dimension $m = m_0 \varepsilon_{rg}^{3/2}$, the measure of mechanical inertia (mass) $m = m_0 \varepsilon_{rg}^{3/2}$, the angular frequencies of the emitted and absorbed radiation $\omega = \omega_0 \varepsilon_{rg}^{-1/2}$ and the durations $t = t_0 \varepsilon_{rg}^{1/2}$.

To find the relative permittivity and permeability expressions dependent on the mass of the body $M$ and the distance from the body, $r$ Wilson and Dicke do not propose a physical-mathematical model (Wilson states in the paper "An Electromagnetic Theory of Gravitation" [7]: "How these variations are produced by the presence of matter in the ether, at distant points, remains to be explained") to deduce the expression of the electromagnetic constants of the vacuum and implicitly the refractive index, but they establish their expression from the condition that they deduce the expression of the gravitational force from the spatial variation of the energy of the systems (which depends on the refractive index) and the deviation of emw in the gravitational field be as in the General Relativity Theory (GRT)

$$n_g = \varepsilon_{gr} = \mu_{gr} = 1 + \frac{2GM}{c_0^2 r}. \tag{1}$$

In paper [1], we hypothesize that a mass in the EMW, $M$, interacts gravitationally with the components of the physical vacuum (particle-antiparticle pairs) determining a Boltzmann-type distribution of the particle density, $N = N_0 \exp(-E_{pg}/kT_v) = N_0 \exp[GMm_v/(kT_v r)] \cong N_0[1 + GMm_v/(kT_v r)]$. The gravitational index is determined, by the polarization of the vacuum, by the additional number of pairs near the mass $M$, $\Delta N_g = N - N_0 \cong N_0[GMm_v/(kT_v r)]$. Using the expression for the relative permittivity for the gas of positronium atoms (we consider that the density of e-p pairs is much higher than the pairs of other particles with higher rest mass) and the relationship between the refractive index and the relative permittivity, we obtain the expression for the gravitational refractive index $n_g = \varepsilon_{rg} = \mu_{rg} = 1 + GMm_{ep}/(kTr)$, for the static case and $\omega \ll \omega_{0ep}$.

To derive these expressions we use the hypothesis that the density of positronium atoms is inversely proportional to the volume of a positronium atom $N_0 \cong 1/(\pi a^3)$. If we approximate the mass of the positronium atom by the identical masses of the e-p pairs $m_{pv} \cong 2m_e$ and the thermal energy corresponding to the e-p pairs is $kT \cong m_e c_0^2$ (de Broglie hypothesis [10]), the refractive index becomes $n_g = \varepsilon_{rg} = (\mu_{rg}) = 1 + 2GM/(c_0^2 r)$. The compatibility of the GRT results with the existence of a gravitational refractive index was demonstrated in the paper "A Covariant Approach to the Gravitational Refractive Index" [2]. We reviewed the physical modeling of the gravitational refractive index in two other papers [3, 4] because the solution in the paper [1] does not consistently apply the Wilson – Dicke (W-D) theory, i.e. the dependence of the parameters of the vacuum components on the internal refractive index of the vacuum.

In the first part of the paper "Optical properties of Vacuum Modelled as e-p Plasma" [3], we model the vacuum as a low-concentration e-p plasma (the fraction of ionized positronium atoms $f_0 \ll 1$) and derive a relationship between the density of positronium atoms, $N_{0v}$, the mass of electrons (positrons) in vacuum, $m_{ev}$, the charge of e-p pairs in vacuum, $e_v^2 = e^2 \varepsilon_{rv}^{-1}$ ($\varepsilon_{rv}$ is the relative internal permittivity of the vacuum, i.e., the internal components of the vacuum/particle-antiparticle pairs interact through fields that have propagation velocity $c_v = c_0 \varepsilon_{rv}^{-1}$) and the natural pulsation of the positronium atom $\omega_{0pv} = 2m_e c_0^2/\hbar$,

$$1 = \frac{m_{ev} \omega_{0pv}^2}{4\pi e_v^2 N_{0v}}, \tag{2}$$

because the relative permittivity is equal to unity and is a dimensionless quantity. This expression is equivalent to the fact that the physical vacuum has external relative permittivity $\varepsilon_{0r} \cong 1$ (i.e., the permittivity measured by an observer in the EMW). It is necessary to introduce the quantities characterizing the particle-antiparticle pairs in the vacuum and the systems they form, ($m_{ev}$, $e_v^2$, etc.) to explain, for example, why the ionization energy of positronium, calculated according to the semiclassical theory (also valid for the components of the vacuum), is equal to the energy of formation of e-p pairs (see section 2.1 of this paper, Eqs. (11), (15)). In the second part of the paper [3] we study the compression of e-p plasma under the action of the gravitational field generated by a mass $M$ and we demonstrate that, in the approximation of a low - concentration plasma, the density of positronium atoms has a

Boltzmann dependence, for a compression of polytropic index $s$,

$$N = N_0 \exp\left(\frac{-E_{pg}}{skT_\upsilon}\right) \cong N_0\left(1 + \frac{-E_{pg}}{skT_\upsilon}\right) = N_0\left(1 + \frac{2GMm_{p\upsilon}}{skT_\upsilon r}\right), \qquad (3)$$

that is, we demonstrate the hypothesis of the change in the density of positronium atoms with mass $m_{p\upsilon}$ by the interaction with mass $M$, from paper [1]. As in paper [1], in deducing the dependence of the particle density on the gravitational field, we do not take into account how the gravitational potential energy $E_{pg}$ and thermal energy $kT$ change for particles inside the vacuum.

In the first part of the paper "The parameters of vacuum modeled as an e-p Plasma" [4], we deduce the temperature of the positronium atom fluid $T_\upsilon$ and the expression for the relative internal permittivity of the vacuum $\varepsilon_{r\upsilon}$ starting from the expression for the external permeability of the vacuum $\varepsilon_{r\upsilon}$ that depends on the vacuum temperature (equivalently, the equality between the external permittivity and external permeability expressed as a function of the internal parameters of the e-p fluid, $\mu_{0r} = \varepsilon_{0r} = 1$). Here we mistakenly considered that positronium atoms in the ortho state have nonzero magnetic moment. Another mistake is that we did not consistently use the results of the W-D papers for the parameters of atoms in a medium with an index modified by the gravitational field and for the parameters of atoms modified by the internal index of the vacuum: $l = l_0 \varepsilon_{r\upsilon}^{-1/2}$, $m = m_0 \varepsilon_{r\upsilon}^{3/2}$, $\omega = \omega_0 \varepsilon_{r\upsilon}^{-1/2}$ and $t = t_0 \varepsilon_{r\upsilon}^{1/2}$. For this reason we obtained an expression for the relative internal permittivity $\varepsilon_{r\upsilon}$ that is not compatible with the mass dependence $m = m_0 \varepsilon_{r\upsilon}^{3/2}$.

The purpose of this paper is to correct the results of modeling the vacuum as a physical medium outlined, within a semiclassical theory, in the works [1, 3, 4]. The return to this problem was prompted by the study of the papers, which address the same problem, in several papers published later by other authors [11, 12, 14 - 17]. The problem of modeling the physical vacuum as a system of electric and magnetic dipoles and deriving the expression for the electric permittivity and magnetic permeability was developed by: Leuchs [11, 12], Urban [14, 15] and Mainland [16, 17].

Leuchs et al. [11, 12] consider the vacuum to be an effective medium with dielectric and paramagnetic properties composed of virtual particle-antiparticle pairs [13]. These pairs arise as a result of the zero-order fluctuations of the vacuum and interact with real particles in the EMW. In the presence of an electrostatic field produced by the EMW systems, the pairs polarize and according to classical and quantum theory the permittivity of the medium becomes $\varepsilon_0 \cong N_0 q_e^2/(m_{0e}\omega_{0e}^2)$ with $N_0 \cong (2m_{0e}c_0/\hbar)^3$ the density of electron-positron (e-p) pairs, $q_e$ the electron charge, $m_{0e}$ rest mass of the electron in the undisturbed vacuum and $\omega_0$ the natural angular frequency of the e-p pairs. In the presence of a magnetostatic field produced by the EMW systems, electrons and positrons move in opposite directions on circular orbits of radii of the order $\hbar/(2m_{0e}c_0)$ generating a nonzero magnetic moment. The electric and magnetic permittivities are equal, $\varepsilon_{0r} = \mu_{0r}$, if $\omega_0 = 2m_{0e}c_0^2/\hbar$.

Urban et al. [14, 15], based on the same hypothesis of the vacuum as a dielectric medium composed of fermion pairs having the average energy $E_j = 2m_{0j}c_0^2$, with the pair density $N_j \cong 2(2m_{0j}c_0/\hbar)^3$, obtain an expression for the permittivity of the vacuum that does not depend on the mass of the pairs $m_{0j}$, $\varepsilon_0 \cong 2q_e^2/(3\hbar c_0)\sum Q_j^2$, $Q_j = q_j/q_e$, $\sum Q_j^2 = 8$. Also, the vacuum behaves as a paramagnetic because the orbital magnetic moment is zero but the spin moment is different from zero, because, in the singlet state, the spin is zero but the charges being of opposite sign, the magnetic moment of the pair is double the magnetic moment of the electron/positron. In the paper [14] the authors also propose a physical-mathematical model of the interaction of the vacuum with the gravitational field through which they deduce the expression for the refractive index of the vacuum produced by the gravitational field. The gravitational field modifies the parameters of the vacuum components, according to the W-D hypothesis, and therefore the lifetime of the virtual pairs, $\tau(r) = \tau_\infty \left(1 + 2GM/(c_0 r)\right)$, which influences the density of the virtual pairs, $N(r) = N_\infty \left(1 + GM/(c_0 r)\right)$, the interaction cross section between photons and virtual pairs and implicitly the gravitational refractive index.

Mainland and Mulligan consider that the vacuum is made up of positronium atoms that are polarized under the action of an electrostatic field [16, 17]. The natural pulsation of the components of the vacuum lepton-antilepton pairs is: $\omega_j = \left|E_{p-Ps}\right|/\hbar = m_j \alpha^2 c_0^2/(4\hbar)$ with $E_{pj} = -(m_j/2) q_e^4 / \left[2(4\pi\varepsilon_0 \hbar)^2\right] = -\alpha^2 m_j c_0^2 / 4$, $\alpha = q_e^2/(4\pi\varepsilon_0 \hbar c_0) = e^2/(\hbar c_0)$. The average duration of the pairs is $\Delta t_j = \hbar/(4 m_j c_0^2)$ and the average distance between the pairs is $L_j = c_0 \Delta t_j = \hbar/(4 m_j c_0)$. With these quantities, the pair density is $N_{L_j} = 1/L_j^3 = (4 m_j c_0)^3/\hbar^3$. The density of paraatoms corresponding to each type of pair is obtained from the pair density corrected with the probability that the paraatom interacts with the incident photon during $\Delta t_j$, $N_j = N_{L_j} \Gamma_j \Delta_j = \left[(4 m_j c_0)^3/\hbar^3\right](\alpha^5/4)$ (with $\Gamma_j = \alpha^5 m_j c_0^2/\hbar$ the electromagnetic decay rate for paraatom after interaction with a photon). With these quantities, the permittivity of the vacuum, for an electrostatic field, has the expression $\varepsilon_0 = \sum \left[2 N_j q_e^2/(m_j \omega_j^2)\right] \cong \sum 8^3 \alpha q_e^2/(\hbar c_0) = 3 \cdot 8^3 \alpha q_e^2/(\hbar c_0)$. Incorrectly, it is considered that paraatoms, in the singlet state, have a totally zero magnetic moment.

The behavior of vacuum as a medium with a refractive index is demonstrated not only for emw but also for matter waves attached to particles with rest mass, i.e. de Broglie waves [18 - 20].

In the second part of the paper we study the optical properties of vacuum: the properties of positronium atoms, the external relative permittivity and permeability of the vacuum, the gravitational refractive index and the special refractive index. In the third part we study the external relative permittivity of the vacuum generated by the ZPF background. In the fourth part we study the imaginary relative permittivity of free vacuum and in the presence of a mass and the attenuation of emw by scattering and absorption. In the fifth part we study the dependence of the external relative index of the vacuum on the angular frequency. In the sixth section we deduce the relative permittivity of the external vacuum under the action of an electrostatic field. In the last section we comment on the results and conclude on the conditions of their validity.

## 2. Optical properties of physical vacuum

### 2.1. Permittivity of free vacuum

In this section we will correct the results obtained in previous papers [1, 3, 4], within a semiclassical model of the vacuum. The main hypothesis is that, analogous to the behavior of particles and atoms in the vacuum modified by the gravitational field, particle-antiparticle pairs and the systems formed by them (paraatoms) interact through the maximum speed $c_v = c_0/\varepsilon_{rv} \gg c_0$. According to the W-D theory, the fundamental parameters change so that the fine constant remains invariant $\alpha_v = e_v^2/(\hbar c_v^2) = e_g^2/(\hbar c_g^2) = e^2/(\hbar c_0^2)$, i.e. $\alpha_v = \alpha_g = \alpha$.

In the following, we will consider that the vacuum is composed of positronium atoms because the probability of formation of pairs with higher rest mass is lower [1, 3, 4, 21- Ch. 12, Sch. 111] and the ionization energy of the positronium atom [15, 22 - Ch. 9, Sch. 84], in the ground state in EMW, is equal to the modulus of the total energy $E_{op}$

$$E_{ip} = \left|E_{op}\right| = \frac{e^4 m_e}{4\hbar^2} = \frac{\alpha^2 m_e c_0^2}{4} \ll m_e c_0^2, \tag{4}$$

The radius of the positronium atom is

$$r_p = \frac{2\hbar^2}{e^2 m_e} = \frac{2\hbar}{\alpha m_e c_0} \tag{5}$$

and the orbital speed is

$$\upsilon_p = \frac{e^2}{\hbar} = c_0 \alpha \ll c_0, \tag{6}$$

so the non-relativistic approximation is good.

The natural pulsation of the positronium atom is

$$\omega_{0p} = \frac{|E_{op}|}{\hbar} = \frac{e^4 m_e}{4\hbar^3} = \frac{\alpha^2 m_e c_0^2}{4\hbar}. \tag{7}$$

The relative permittivity of the system of positronium atoms, with density in a variable electric field, is [23, Ch. 31]

$$\varepsilon_r = 1 + \frac{Nq_e^2}{2\varepsilon_0 m_e \left(\omega_{0p}^2 - \omega^2 - i\omega\gamma_t\right)} = 1 + \frac{Nq_e^2 \left(\omega_{0p}^2 - \omega^2 + i\omega\gamma_t\right)}{2\varepsilon_0 m_e \left[\left(\omega_{0p}^2 - \omega^2\right)^2 + \omega^2\gamma_t^2\right]} = $$
$$1 + \frac{Nq_e^2 \left(\omega_{0p}^2 - \omega^2\right)}{2\varepsilon_0 m_e \left[\left(\omega_{0p}^2 - \omega^2\right)^2 + \omega^2\gamma_t^2\right]} + \frac{iNq_e^2 \omega\gamma_t}{2\varepsilon_0 m_e \left[\left(\omega_{0p}^2 - \omega^2\right)^2 + \omega^2\gamma_t^2\right]}, \tag{8}$$

with $\gamma_t = \gamma' + 2\omega^2 e^2/(3m_e c_0^3)$ the total decay constant and the decay constant of the real dissipation phenomena (absorption) and $2\omega^2 e^2/(3m_e c_0^3)$ the decay constant of the scattering dissipation phenomena [24, Ch. 17]. The real part is the permittivity of the refraction phenomena and the imaginary part characterizes the absorption phenomena [25, Ch. 32]. The real permittivity, for low frequencies (electrostatic fields, i.e. $\omega \to 0$), becomes

$$\varepsilon_{Rr} = \lim_{\omega \to 0}\left\{1 + \frac{Nq_e^2 \left(\omega_{0p}^2 - \omega^2\right)}{2\varepsilon_0 m_e \left[\left(\omega_{0p}^2 - \omega^2\right)^2 + \omega^2\gamma_t^2\right]}\right\} = 1 + \frac{Nq_e^2}{2\varepsilon_0 m_e \omega_{0p}^2} = 1 + \frac{2\pi Ne^2}{m_e \omega_{0p}^2}. \tag{9}$$

Replacing the natural pulsation, according to Eq. (7), we get

$$\varepsilon_{Rr} = 1 + \frac{2^5 \pi N \hbar^6}{m_e^3 e^6} = 1 + \frac{2^5 \pi N \hbar^3}{\alpha^3 m_e^3 c_0^3}. \tag{10}$$

For positronium atoms inside the physical vacuum, characterized by the propagation speed of interactions $c_\upsilon = c_0/\varepsilon_{r\upsilon}$, characteristic parameters are: ionization energy

$$E_{ip\upsilon} = |E_{op\upsilon}| = \frac{e_\upsilon^4 m_{e\upsilon}}{4\hbar^2} = \frac{\alpha^2 m_{e\upsilon} c_\upsilon^2}{4}, \tag{11}$$

natural angular frequency

$$\omega_{0p\upsilon} = \omega_{0\upsilon} = \frac{|E_{op\upsilon}|}{\hbar} = \frac{e_\upsilon^4 m_{e\upsilon}}{4\hbar^3}, \tag{12}$$

radius

$$r_{p\upsilon} = \frac{2\hbar^2}{e_\upsilon^2 m_{e\upsilon}} = \frac{2\hbar}{\alpha m_{e\upsilon} c_\upsilon} \tag{13}$$

and orbital speed

$$\upsilon_{p\upsilon} = \frac{e_\upsilon^2}{\hbar} = \frac{c_0 \alpha e_\upsilon^2}{e^2}. \tag{14}$$

To extract the e-p pair from the vacuum, a minimum energy approximately equal to the mass of the positronium atom, $E_{\min} \cong m_{ap}c_0^2 \cong 2m_e c_0^2 = E_{ip} = \hbar \omega_{0\upsilon}$, is required. From this equality and the expression for the mass of the electron in vacuum $m_{e\upsilon} = m_e \varepsilon_{r\upsilon}^{3/2}$, the expression for the relative internal permittivity of the vacuum results

$$2m_e c_0^2 = \frac{e_\upsilon^4 m_{e\upsilon}}{4\hbar^2} \text{ or } \frac{e_\upsilon^4 m_{e\upsilon}}{8\hbar^2 m_e c_0^2} = \frac{e^4}{8\varepsilon_{r\upsilon}^{1/2} \hbar^2 c_0^2} = 1 \text{ or } \varepsilon_{r\upsilon} = \left(\frac{e^2}{\hbar c_0}\right)^4 \frac{1}{8^2} = \frac{\alpha^4}{2^6} \ll 1. \tag{15}$$

With this expression of the internal relative permittivity, the parameters of the positronium atom inside the vacuum are: natural angular frequency

$$\omega_{0p\upsilon} = \omega_{0\upsilon} = \frac{e_\upsilon^4 m_{e\upsilon}}{4\hbar^3} = \frac{2m_e c_0^2}{\hbar}, \tag{16}$$

radius

$$r_{p\upsilon} = \frac{2\hbar}{\alpha m_{e\upsilon} c_\upsilon} = \frac{2^4 \hbar}{\alpha^3 m_e c_0} \gg r_p = \frac{2\hbar}{\alpha m_e c_0} \gg \frac{\hbar}{m_e c_0} \tag{17}$$

and orbital speed

$$\upsilon_{p\upsilon} = \frac{e_\upsilon^2}{\hbar} = \frac{2^6 c_0}{\alpha^3} \gg \upsilon_p = c_0 \alpha \text{ or } \upsilon_{p\upsilon} = \frac{e_\upsilon^2}{\hbar} = \frac{c_0}{\varepsilon_{r\upsilon}} \frac{e^2}{\hbar c_0} = c_\upsilon \alpha \ll c_\upsilon. \tag{18}$$

It follows that the non-relativistic approximation is also good for the positronium atom inside the vacuum.

If we use the expression of relative permittivity for real atoms given by Eq. (9), multiplied by 2 (for vacuum, the refractive index is equal to the relative permittivity and not to a radical of the permittivity), the expression of the external relative permittivity of the vacuum (for electrostatic fields, i.e. $\omega \to 0$), is determined by the positronium atoms in the vacuum with density $N_{0\upsilon}$, relative to the EMW reference system, and the pulsation given by (16)

$$\varepsilon_{0r} = \frac{4\pi N_{0\upsilon} e_\upsilon^2}{m_{e\upsilon} \omega_{0\upsilon}^2} = \frac{\pi N_{0\upsilon} e_\upsilon^2}{m_{e\upsilon} m_e^2 c_0^4}. \tag{19}$$

The same permittivity can also be expressed in terms of the eigenparameters of positronium in vacuum (i.e.: $N_0, m_e, e^2$ and $m_{e\upsilon} = m_e \varepsilon_{r\upsilon}^{3/2}$, $e_\upsilon^2 = e^2 \varepsilon_{r\upsilon}^{-1}$, $N_{o\upsilon} = N_0 \varepsilon_{r\upsilon}^{3/2}$, because $N_{0\upsilon} = \mathbb{N}/V_\upsilon = \mathbb{N}/(V_0 \varepsilon_{r\upsilon}^{-3/2}) = N_0 \varepsilon_{r\upsilon}^{3/2}$)

$$\varepsilon_{0r} = \frac{\pi N_0 e^2 \hbar^2}{m_e^3 c_0^4 \varepsilon_{r\upsilon}}. \tag{20}$$

This expression differs from the one used by: Simaciu [1, 3, 4], Leuchs [11, 12], Urban [14, 15] Mainland [16, 17] and collaborators, $\varepsilon_{0r} = 2\pi N_0 e^2 \hbar^2 / (m_e \omega_{0\upsilon}^2) = \pi N_0 e^2 \hbar^2 / (2m_e^3 c_0^4)$ in which only $\omega_{0\upsilon} = 2m_e c_0^2 / \hbar$. To establish the correct expression, regardless of whether we express it in terms of proper and relative parameters [26], we return to the expression of the external vacuum permittivity

$$\varepsilon_0 \varepsilon_{0r} = \frac{N_{0\upsilon} q_e^2}{m_{e\upsilon} \omega_{0\upsilon}^2} = \frac{N_{0\upsilon} q_e^2 \hbar^2}{4 m_{e\upsilon} m_e^2 c_0^4} = \frac{\pi N_0 \varepsilon_0 e^2 \hbar^2}{m_e^3 c_0^4} \tag{21}$$

and so the expression for the external relative permittivity of the vacuum is

$$\varepsilon_{0r} = \frac{\pi N_0 e^2 \hbar^2}{m_e^3 c_0^4}. \tag{22}$$

In the singlet state of positronium atoms (parapositronium), the orbital magnetic moment is zero, since the charges are opposite and the pairs rotate in the same direction. The spin magnetic moment is the magnetic moment of the fermion if the pairs have zero spin moment but opposite charges [12]. The positronium fluid behaves like a paramagnet in a stationary magnetic field and the relative permittivity [25, Ch. 34] has the expression:

$$\mu_r = \frac{\mu_0 g^2 j(j+1) N \mu_B^2}{12kT}, \tag{23}$$

with $T$ the temperature of the positronium fluid with respect to the EMW. Since only parapositronium atoms participate in the magnetization phenomenon (which are mixed, with almost equal probabilities, with orthopositronium atoms in the three states that have zero spin magnetic moment $N_{pp} \cong N_{op+} \cong N_{op-} \cong N_{op0} = N/4$ [56]), in expression (23) we replaced the particle density $N$ with $N \to N/4$. Replacing: $g = 2$, $j = 1/2$, and the Bohr magneton $\mu_B = q_e \hbar / (2m_e)$ in (23), we obtain

$$\mu_r = \frac{N \mu_0 q_e^2 \hbar^2}{4kT m_e^2} = \frac{\pi N e^2 \hbar^2}{kT m_e^2 c_0^2}. \tag{24}$$

For the positronium fluid in vacuum, with density and temperature, measured with vacuum etalons [26], the expression for the relative permittivity is

$$\mu_{0r} = \frac{N_0 \mu_0 q_e^2 \hbar^2}{4kT_0 m_e^2} = \frac{\pi N_0 e^2 \hbar^2}{kT_0 m_e^2 c_0^2}. \tag{25}$$

The same relative permeability can also be expressed as a function of the parameters of the positronium in the vacuum, measured with the etalons outside the vacuum

$$\mu_{0r} = \frac{\pi N_{0\upsilon} e_\upsilon^2 \hbar^2}{kT_{0\upsilon} m_{e\upsilon}^2 c_\upsilon^2}. \tag{26}$$

We specify that both from the relativistic point of view (special and general) and from the interpretation of relativistic effects as effects of the modification of the parameters of objects, phenomena and measuring devices (implicitly of the etalons) [26], the physical quantities noted: $N_0$, $T_0$, $l_0$, $c_0$, $m_e$, $\omega_0$ are proper quantities (quantities from the moving reference system or in the gravitational field measured with the etalons from the moving system or in the field) and the quantities noted in the same way but with the index attached below $\upsilon$, $g$ or $V$ (for example: $N_{0\upsilon}$, $T_{0\upsilon}$, $N_{gV}$ etc.) are mixed/relative quantities (quantities from the moving reference system or in the gravitational field measured with the etalons from the system at rest or from outside the field).

From the equality of permittivity (22) and relative permeability (25), $\mu_{0r} = \varepsilon_{0r}$, results in the average thermal energy of the vacuum components (positronium atoms, positrons and free electrons)

$$kT_0 = m_e c_0^2. \tag{27}$$

This relationship is compatible with the hypothesis of L. de Broglie [10].

If $\varepsilon_{0r} = \varepsilon_{0Rr} = 1$, then from Eq. (22), the expression for the density of positronium atoms results

$$N_0 = \frac{m_e^3 c_0^4}{\pi e^2 \hbar^2} = \frac{1}{\pi r_e \lambda_C^2}, \tag{28}$$

which differs from the expressions obtained in the works [11, 12, 13 - 17] proportional to $N_0 \sim 1/\lambda_C^3$. Substituting in Eq. (28) the values of the physical quantities, it results that the density of positronium atoms is $N_0 \cong 10^{40}\,\text{m}^{-3}$. Particle density measured with etalons outside the vacuum (EMW) has the value $N_{0\upsilon} = N_0 \varepsilon_{r\upsilon}^{3/2} = N_0 \left(\alpha^6/2^9\right) \cong 10^{24}\,\text{m}^{-3}$.

In the papers of Leuchs and collaborators, for e-p pairs $N_{0e} \cong 10^{39}\,\text{m}^{-3}$. In the papers of Urban and collaborators, for e-p pairs $N_{0e} \cong 2.8\,10^{38}\,\text{m}^{-3}$. In the papers of Mainland and Mulligan, for e-p pairs $N_{0e} \cong 1.12\,10^{39}\,\text{m}^{-3}$. The expression for the external relative permittivity of the vacuum includes this particle density corrected with the probability that a parapositronium in the vacuum absorbs a photon during a time $\Delta t = 4\hbar/\left(m_e c_0^2\right)$, $N_{P-Ps} = N_{0e}\Gamma_{P-Ps}\Delta t \cong 10^{28}\,\text{m}^{-3}$ and the angular frequency given by Eq. (7).

## 2.2. Gravitational and Special Refractive Index

To derive the expression for the gravitational refractive index and the special refractive index (the index dependent on the relative velocity [27]), we return to the expression for the density of the fluid of positronium atoms compressed under the action of the gravitational field (1), derived in the paper [1]. To be consistent with the W-D hypothesis, the expression for the gravitational potential energy of interaction between the mass $M$ outside the vacuum and the mass $m_{p\upsilon}$ of the positronium atom in the vacuum must be rewritten, because in relation (1) we did not take into account that both the temperature and the gravitational constant in the vacuum are dependent on the relative internal permittivity of the vacuum $\varepsilon_{r\upsilon}$.

We derive the dependence of the gravitational constant on the gravitational permittivity by introducing the notion of gravitational charge $e_g = \sqrt{G}m$. With this, the potential energy between two identical masses is

$$E_{pg} = \frac{-Gm^2}{r} = \frac{-e_g^2}{r}, \tag{29}$$

If both masses are immersed in a medium with relative permittivity $\varepsilon_r$, which causes the change in mass, $m_r = m\varepsilon_r^{3/2}$, distance $r_r = r\varepsilon_r^{-1/2}$ and gravitational constant, $G_r$, then the expression for potential energy becomes

$$E_{pgr} = \frac{-G_r m_r^2}{r_r} = \frac{-G_r m^2 \varepsilon_r^3}{r\varepsilon_r^{-1/2}} \equiv E_{pg}\varepsilon_r^{-1/2} = \frac{-Gm^2 \varepsilon_r^{-1/2}}{r}, \tag{30}$$

that is, the gravitational interaction constant in the medium, depends on the gravitational constant through the relationship

$$G_r = G\varepsilon_r^{-4}. \tag{31}$$

These dependencies are also obtained in the theory of Peter Rastall [26, 28, 29].

According to these dependencies, the gravitational interaction potential energy between the gravitational charge $e_{gM} = \sqrt{G}M$ and the gravitational charge $e_{gm} = \sqrt{G_r}m_r = \sqrt{G}m_r\varepsilon_r^{-2}$, at the distance $r$, has the expression

$$E_{pgr} = \frac{-GMm_r\varepsilon_r^{-2}}{r}. \tag{32}$$

If the medium is the positronium fluid: $\varepsilon_r \to \varepsilon_{r\upsilon}$ and $E_{pg\upsilon} = \left(-GMm_\upsilon \varepsilon_{r\upsilon}^{-2}\right)/r = \left(-GMm\varepsilon_{r\upsilon}^{-1/2}\right)/r = E_{pg}\varepsilon_{r\upsilon}^{-1/2}$.

To derive the gravitational refractive index and the special refractive index we consider that the mass $M$ moves with velocity $V$ relative to a reference frame at rest (i.e. for which the equilibrium thermal radiation is isotropic [27]). For a system bound to the mass $M$, the positronium atoms in the vacuum also have a translational motion with velocity and hence a kinetic energy $E_{kV} = m_a V^2/2$. These atoms in ordered motion which are also in the gravitational field of the mass $M$ have a total energy $E_t = E_p + E_k$ and obey a Boltzmann distribution $N = N_0 \exp\left[-E_p/(kT_0) - E_k/(kT_0)\right]$ or

$$N_{gV} \cong N_0 \left(1 + \frac{GM m_a}{kT_0 r} - \frac{m_a V^2}{2kT_0}\right). \tag{33}$$

Since the ratio of the total energy to the Boltzmann energy $kT_0$ is a number, this ratio remains unchanged if we write it with the physical quantities in the vacuum

$$N_{gV} \cong N_0 \left(1 + \frac{\sqrt{G}M\sqrt{G_\upsilon} m_{ap\upsilon}}{kT_{0\upsilon} r} - \frac{m_{ap\upsilon} V_\upsilon^2}{2kT_{0\upsilon}}\right) = N_0 \left(1 + \frac{\sqrt{G_\upsilon} M_\upsilon \sqrt{G_\upsilon} m_{ap\upsilon}}{kT_{0\upsilon} r_\upsilon} - \frac{m_{ap\upsilon} V_\upsilon^2}{2kT_{0\upsilon}}\right) = \\ N_0 \left(1 + \frac{GM m_{ap}}{kT_0 r} - \frac{m_{ap} V^2}{2kT_0}\right). \tag{34}$$

In the deriving this expression we have taken into account that the Boltzmann constant is a relativistic invariant, $k_\upsilon = k$, and the temperature transforms according to the relation $T_{0\upsilon} = T_0 \varepsilon_{r\upsilon}^{-1/2}$ by analogy with the transformation relation within the GRT [30]. Replacing also the Boltzmann energy expression given by Eq. (27), with $m_{ap} \cong 2m_e$, we obtain

$$N_{gV} \cong N_0 \left(1 + \frac{2GM}{c_0^2 r} - \frac{V^2}{c_0^2}\right). \tag{35}$$

Substituting expression (35) into the relative permittivity expression (22), with $N_0 \to N_{gV}$ and $\varepsilon_{0r} = 1$, we obtain

$$n_{gV} = \varepsilon_{0gVr} = \frac{\pi N_{gV} e^2 \hbar^2}{m_e^3 c_0^4} = \frac{\pi N_0 e^2 \hbar^2}{m_e^3 c_0^4}\left(1 + \frac{2GM}{c_0^2 r} - \frac{V^2}{c_0^2}\right) = 1 + \frac{2GM}{c_0^2 r} - \frac{V^2}{c_0^2}. \tag{36}$$

With this expression of permittivity and refractive index, according to W-D theory, the length dependence in a medium with permittivity given by (36) is

$$l = l_0 \varepsilon_{0gVr}^{-1/2} = l_0 \left(1 + \frac{2GM}{c_0^2 r} - \frac{V^2}{c_0^2}\right)^{-1/2}. \tag{37}$$

If the mass $M$ is at rest ($V = 0$), the external refractive index of the vacuum around it has the expression given by Eq. (1), i.e. the gravitational index and the dependence of lengths on it $l_g = l_0 \left[1 - GM/(c_0^2 r)\right] < l_0$.

If the mass is zero, $M = 0$, the external refractive index of the vacuum relative to a moving observer is

$$n_V = 1 - \frac{V^2}{c_0^2}, \tag{38}$$

and $l_V = l_{0V}\left[1 + V^2/(2c_0^2)\right] > l_{0V}$. This dependence of lengths is not similar to that given by the Lorentz transformations ($l = l_0 \sqrt{1 - V^2/c_0^2} < l_0$). The inconsistency results from the fact that, in order to deduce the expression of the external permittivity of the vacuum, we considered that the positronium atoms move relative to the system attached to the mass $M$, although we initially considered that the mass $M$ is in motion. According to the definition of relativistic effects as effects of the measurement process combined with the real modification of the parameters of objects, phenomena and physical etalons [26], we must redefine the type of length (proper length or mixed/relative length). In the dependence relation $l_v = l_{0v}\left[1+V^2/(2c_0^2)\right]$, $l_{0v} \to l_v$, that is the relative length and $l_v \to l_{0v} = l_0$ is the proper length: $l_0 = l_v\left[1+V^2/(2c_0^2)\right]$ or $l_v = l_0\left[1-V^2/(2c_0^2)\right] \cong l_0\sqrt{1-V^2/c_0^2}$.

According to the expression (35) of the relative permittivity induced by a moving mass, the effects of special relativity appear as corrections of the effects of gravitational refraction, but they do not give a refraction effect themselves [27]. To obtain the expression of the special permittivity it is necessary to introduce an absolute system, such as the one related to $2.7\,\text{K}$ thermal radiation, according to the method in the paper "Generalized Tensor of Relativistic Permittivity" [27]

## 3. External relative permittivity of the vacuum and the ZPF background

Analogous to the structure and phenomena occurring in the Acoustic World (AW) [31, 32], the physical vacuum (in the EMW) has two levels of structure: a fundamental level corresponding to the fluid/ liquid in the AW and a secondary level determined by the background Zero-Point Field or Classical Zero-Point Field (ZPF or Zero-Point Energy – ZPE, CZPF [33-40]- i.e. the primary electromagnetic radiation scattered by real and virtual particles) and the gravitational potential generated by the particles in the Universe. According to Mach's Principle and the W-D hypothesis [7, 8], the gravitational potential generated by the particles in the Universe is

$$\Phi_U = \int_0^R \frac{2G\rho_U dV}{r} = c_0^2. \tag{39}$$

The particles in the Universe (i.e. real electrons and protons in EMW) interact both with the secondary electromagnetic radiation that forms the electromagnetic field, around electric charges in translational and orbital motions, whose energy density varies as $w_{em} \sim 1/r^p$, $p \geq 3$ well as with the thermal background $2.7\,\text{K}$ and the ZPF/CZPF background as an effect of the internal motions of the non-point-modeled particles [38-40].

The relative permittivity induced by the ZPF/ZPE or CZPF/CZPE can be calculated using the expression of the permittivity tensor generated by an electromagnetic field obtained byc and Kokel [41, 24 - Linear Superposition], recalculated in the papers [42, 43] for a stochastic electromagnetic wave background,

$$\varepsilon_{0rZPF} = \frac{28e^4\hbar w_{CZPF}}{135m_e^4c_0^7} = \frac{14e^4\hbar^2}{135\pi^2 m_e^4 c_0^{10}}\int_0^{\omega_l}\omega^3 d\omega = \frac{7e^4\hbar^2\omega_l^4}{270\pi^2 m_e^4 c_0^{10}}. \tag{40}$$

This expression of the relative permittivity allows us to find the limiting angular frequency if we consider the relative permittivity to be equal to unity. From the condition $\varepsilon_{0CZPF}=1$, it follows that $\omega_l = \left(2m_e c_0^2/\hbar\right)\left[(135/56)^{1/4}\left(\hbar c_0/e^2\right)^{1/2}\right] > 2m_e c_0^2/\hbar$, but smaller than the angular frequency corresponding to electrons and positrons in vacuum $\omega_{ev} = \left(m_e c_0^2/\hbar\right)\varepsilon_{rv}^{-1/2} = \left(m_e c_0^2/\hbar\right)\left(2^3 \hbar c_0/e^2\right)^2$.

To avoid imposing a limit on the CZPF spectrum, we can use a property of the ZPF spectrum that it is perceived by an accelerated particle as a Planck spectrum with the temperature given by the Fulling-Unruh-Devis formula [44 - 47]

$$T = \frac{\hbar a}{2\pi c_0 k}. \tag{41}$$

For the electron at rest, with rest energy $\hbar\omega_0/2 = m_e c_0^2$, the semiclassical Barut–Zanghi model [48-50] (i.e., a charge in inertial motion with velocity $c_0$, on an orbit of radius $r_0 = c_0/\omega_0 = \hbar/(2m_e c_0)$) predicts a radial acceleration $a_e = c_0^2/r_0 = 2m_e c_0^3/\hbar$ and hence the corresponding temperature is

$$T_e = \frac{\hbar a_e}{2\pi c_0 k} = \frac{m_e c_0^2}{\pi k} \text{ or } kT_e = \frac{m_e c_0^2}{\pi}. \tag{42}$$

This temperature expression is similar, up to a small constant, to the temperature attached to the electron at translational rest, in the L. de Broglie model [10, 50] or to Eq. (27) obtained in section 2.1. With this temperature expression given by Eq. (42), we calculate the Planckian background density with temperature $T_e$

$$w(T_e) = \int_0^\infty \frac{\hbar \omega^3 d\omega}{\pi^2 c_0^3 \left[\exp(\hbar\omega/kT_e) - 1\right]} = \frac{6(1.08) k^4 T_e^4}{\pi^2 c_0^3 \hbar^3} = \frac{6(1.08) m_e^4 c_0^5}{\pi^6 \hbar^3}. \tag{43}$$

We obtain the expression for the relative permittivity by replacing $w_{CZPF}$ with $w(T_e)$ in Eq. (40)

$$\varepsilon_{0T_e} = \frac{28 e^4 \hbar w(T_e)}{135 m_e^4 c_0^7} = \frac{168(1.08)}{135 \pi^6} \left(\frac{e^2}{\hbar c_0}\right)^2 \ll 1. \tag{44}$$

It follows that the vacuum model based on the positronium atom fluid (i.e. limiting the model to only the contribution of the lightest charged fermions) leads to the inference of a negligible contribution of the ZPF photon background to the relative permittivity.

Under these conditions, the relative external permittivity of the vacuum is determined either by the positronium atom fluid or by the gravitational field determined by all the masses in the finite model of the universe. Since, according to the W-D hypothesis, the gravitational interaction is of electromagnetic nature (the effect of vacuum polarization), the two permittivities must necessary to coincide. It remains to demonstrate this coincidence ($\varepsilon_{0rt} = \varepsilon_{0rg} = \varepsilon_{0r}$) directly in a future paper.

## 4. Relative imaginary permittivity of vacuum

### 4.1. The vacuum away from the masses

For time-varying electromagnetic fields, we can also define an imaginary relative permittivity of the vacuum [1, 25]. According to Eq. (8), the refractive index corresponding to this permittivity has the expression

$$n_I = \varepsilon_{Ir} = \frac{N_0 q_e^2 \omega \gamma_{pt}}{\varepsilon_0 m_e \left[\left(\omega_{0p}^2 - \omega^2\right)^2 + \omega^2 \gamma_{pt}^2\right]} = \frac{4\pi N_0 e^2 \omega \gamma_{pt}}{m_e \left[\left(\omega_{0p}^2 - \omega^2\right)^2 + \omega^2 \gamma_{pt}^2\right]}. \tag{45}$$

We assumed that for the relative magnetic permeability of vacuum we can define a dependence on the angular frequencies $\omega$ and $\omega_{0p}$ [51] and therefore, for vacuum $n_I = \sqrt{\varepsilon_{Ir}\mu_{Ir}} = \varepsilon_{Ir}$, if $\mu_{Ir} = \varepsilon_{Ir}$. Substituting in Eq. (45), the expression $\gamma_{pt} = \gamma'_p + 2\omega^2 e^2/(3m_e c_0^3)$ and taking into account, according to relation (19), that $4\pi N_0 e^2/(m_e \omega_{0\upsilon}^2) = 1$ ($\omega_{0p} = \omega_{0\upsilon}$), we obtain

$$n_I = \varepsilon_{0Ir} = \frac{\omega_{0p}^2 \left[\gamma'_p + 2\omega^2 e^2/(3m_e c_0^3)\right]\omega}{\left[\left(\omega_{0p}^2 - \omega^2\right)^2 + \omega^2 \gamma_{pt}^2\right]} = \frac{\omega_{0p}^2 \gamma'_p \omega}{\left[\left(\omega_{0p}^2 - \omega^2\right)^2 + \omega^2 \gamma_{pt}^2\right]} +$$

$$\frac{2\omega_{0p}^2 e^2 \omega^3}{3 m_e c_0^3 \left[\left(\omega_{0p}^2 - \omega^2\right)^2 + \omega^2 \gamma_{pt}^2\right]} = \varepsilon'_{0Ir} + \varepsilon_{0Irsc}, \tag{46}$$

It follows that, for vacuum we distinguish an imaginary permittivity determined by the absorption of EM waves, $\varepsilon'_{Ir}$ and an imaginary permittivity determined by the scattering of EM waves, $\varepsilon_{Irsc}$. These imaginary permittivities correspond to an attenuation coefficient and an attenuation distance [25- Ch. 32], according to the relations

$$\beta = \frac{2\omega\varepsilon_I}{c_0}, \quad D = \frac{1}{\beta} = \frac{c_0}{2\omega\varepsilon_I}. \tag{47}$$

For the vacuum far from the masses, with $\hbar\omega_{0\upsilon} = 2m_e c_0^2$, it results: a) for the phenomenon of attenuation by absorption

$$\beta' = \frac{8\omega^2 m_e^2 c_0^3 \gamma'_{pt}}{\hbar^2\left[\left(\omega_{0p}^2 - \omega^2\right)^2 + \omega^2\gamma_{pt}^2\right]}, \quad D' = \frac{\hbar^2\left[\left(\omega_{0p}^2 - \omega^2\right)^2 + \omega^2\gamma_{pt}^2\right]}{8\omega^2 m_e^2 c_0^3 \gamma'_{pt}} \tag{48}$$

and b) for the phenomenon of scattering attenuation

$$\beta_{sc} = \frac{16e^2 m_e \omega^4}{3\hbar^2\left[\left(\omega_{0p}^2 - \omega^2\right)^2 + \omega^2\gamma_{pt}^2\right]}, \quad D_{sc} = \frac{3\hbar^2\left[\left(\omega_{0p}^2 - \omega^2\right)^2 + \omega^2\gamma_{pt}^2\right]}{16e^2 m_e \omega^4}. \tag{49}$$

Since $\gamma'_p \ll 2\omega^2 e^2/(3m_e c_0^3)$, it follows that $\gamma_{pt} \cong 2\omega^2 e^2/(3m_e c_0^3)$ and we can approximate expressions (48, 49) in the form

$$\beta' \cong \frac{8\omega^2 m_e^2 c_0^3 \gamma'_p}{\hbar^2\left[\left(\omega_{0p}^2 - \omega^2\right)^2 + 4\omega^6 e^4/\left(9m_e^2 c_0^6\right)\right]}, \quad D' \cong \frac{\hbar^2\left[\left(\omega_{0p}^2 - \omega^2\right)^2 + 4\omega^6 e^4/\left(9m_e^2 c_0^6\right)\right]}{8\omega^2 m_e^2 c_0^3 \gamma'_p}. \tag{50}$$

$$\beta_{sc} = \frac{16e^2 m_e \omega^4}{3\hbar^2\left[\left(\omega_{0p}^2 - \omega^2\right)^2 + 4\omega^6 e^4/\left(9m_e^2 c_0^6\right)\right]}, \quad D_{sc} = \frac{3\hbar^2\left[\left(\omega_{0p}^2 - \omega^2\right)^2 + 4\omega^6 e^4/\left(9m_e^2 c_0^6\right)\right]}{16e^2 m_e \omega^4}. \tag{51}$$

For small angular frequencies, $\omega \ll \omega_{0\upsilon} = 2m_e c_0^2/\hbar$, these quantities become:

$$\beta' \cong \frac{8\omega^2 m_e^2 c_0^3 \gamma'_p}{\hbar^2 \omega_{0p}^4} = \frac{\hbar^2 \omega^2 \gamma'_p}{2m_e^2 c_0^5}, \quad D' \cong \frac{2m_e^2 c_0^5}{\hbar^2 \omega^2 \gamma'_p}, \quad \lim_{\omega\to 0} D' = \infty; \tag{52}$$

$$\beta_{sc} \cong \frac{16e^2 m_e \omega^4}{3\hbar^2 \omega_{0p}^4} = \frac{e^2 \hbar^2 \omega^4}{3m_e^3 c_0^8}, \quad D_{sc} \cong \frac{3m_e^3 c_0^8}{e^2 \hbar^2 \omega^4}, \quad \lim_{\omega\to 0} D_{sc} = \infty. \tag{53}$$

For frequencies close to the angular resonance frequency, $\omega \cong \omega_{0\upsilon} = 2m_e c_0^2/\hbar$, these quantities become:

$$\beta'_0 \cong \frac{18m_e^4 c_0^9 \gamma'_{p0}}{\hbar^2 e^4 \omega_{0p}^4} = \left(\frac{\hbar c_0}{e^2}\right)^2 \frac{9\gamma'_{p0}}{8c_0}, \quad D'_0 \cong \left(\frac{e^2}{\hbar c_0}\right)^2 \frac{8c_0}{9\gamma'_{p0}}; \tag{54}$$

$$\beta_{0sc} \cong \frac{12m_e^3 c_0^6}{\hbar^2 e^2 \omega_{0p}^2} = \frac{3m_e c_0^2}{e^2}, \quad D_{0sc} \cong \frac{e^2}{3m_e c_0^2} = \frac{r_e}{3}. \tag{55}$$

If, according to the paper [38], $\gamma'_{p0} = 2\omega_{0\upsilon}^2 r_{ge}/(3c_0) = 8m_e^2 c_0^3 r_{ge}/(3\hbar^2)$, the gravitational radius $r_{ge} = Gm_e/c_0^2$ and the Schwarzschild radius, $r_{geS} = 2r_{ge}$, substituting in Eq. (54), it follows that the absorption attenuation distance (inside the vacuum) is $D'_0 \cong \alpha^2 \hbar^2 c_0/(3m_e^2 c_0^2 r_{ge}) = r_e^2/(3r_{ge}) \cong 10^{27}$ m, i.e. a distance of the order of the radius of the universe (i.e., EMW) at present, $R_U = GM_U/c_0^2 = N_U(Gm_{pr}/c_0^2)$ ($m_{pr}$ is the proton mass). The scattering attenuation distance, in EMW [52, 53], is of the same order of magnitude ($D_{sc} = 1/(N_U \sigma_{Te}) \cong R_U^3/(N_U r_e^2) \cong R_U^2 r_{gp}/r_e^2 \cong R_U = 10^{26}$ m since $\sigma_{Te} = 8\pi r_e^2/3$, $R_U = N_U r_{gp}$ and $R_U r_{gpr}/r_e^2 \cong 1$, $r_{gp} = Gm_{pr}/c_0^2$).

The ZPF/CZPF background, although attenuated by scattering, is restored, as a spectral energy density, so that the spectral density $\rho(\omega) = \hbar\omega^3/(2\pi^2 c_0^3)$ is the same at every point of the attenuation sphere [54, 55].

According to Eq. (55), the scattering attenuation distance of angular frequencies close in magnitude to $\omega_{0\upsilon} = 2m_e c_0^2/\hbar$, determined by the vacuum components, is of the order of the classical electron radius $D_{0sc} \cong r_e/3 \cong 10^{-15}$ m. This limitation implies that, inside the vacuum, only the particle-antiparticle pairs in the volume $D_{0sc}^3 = r_e^3/27$ interact directly electromagnetically through the angular frequencies $\omega_{0\upsilon}$. Since, according to Eq. (17), the radius of the positronium atom in vacuum is $r_{p\upsilon} = 2^4 \hbar/(\alpha^3 m_e c_0) = 2^4 r_e/\alpha^4 \gg r_e \cong D_{0sc}$, it follows that, in practice, a very small number of pairs interact directly with each other at the frequencies $\omega_{0\upsilon}$. For large angular frequencies, $\omega \gg \omega_{0\upsilon} = 2m_e c_0^2/\hbar$, these quantities become:

$$\beta' \cong \left(\frac{\hbar c_0}{e^2}\right)^2 \left(\frac{m_e c_0^2}{\hbar\omega}\right)^4 \frac{18\gamma'_p}{c_0} \to 0, \quad D' \cong \left(\frac{e^2}{\hbar c_0}\right)^2 \left(\frac{\hbar\omega}{m_e c_0^2}\right)^4 \frac{c_0}{18\gamma'_p}, \lim_{\omega\to\infty} D' = \infty. \tag{56}$$

$$\beta_{sc} \cong \left(\frac{m_e c_0^2}{\hbar\omega}\right)^2 \frac{12 m_e c_0^2}{e^2} \to 0, \quad D_{sc} \cong \left(\frac{\hbar\omega}{m_e c_0^2}\right)^2 \frac{e^2}{12 m_e c_0^2} \to \infty. \tag{57}$$

It follows that the absorption phenomenon is small for frequencies much lower or much higher than the frequency $\omega_{0\upsilon}$.

### 4.2. Vacuum near a body with mass

In interaction with a body of mass $M$, the imaginary relative permittivity of vacuum [1] is

$$n_I = \varepsilon_{Ir} = \frac{4\pi N_0 \left(1 + 2GM/(c_0^2 r)\right) e^2 \omega \gamma_{pt}}{m_e\left[\left(\omega_{0p}^2 - \omega^2\right)^2 + \omega^2\gamma_{pt}^2\right]} = \frac{4\pi N_0 e^2 \omega \gamma_{pt}}{m_e\left[\left(\omega_{0p}^2 - \omega^2\right)^2 + \omega^2\gamma_{pt}^2\right]} + \frac{8\pi GM N_0 e^2 \omega \gamma_{pt}}{m_e c_0^2 r\left[\left(\omega_{0p}^2 - \omega^2\right)^2 + \omega^2\gamma_{pt}^2\right]} = \varepsilon_{0Ir} + \varepsilon_{0gIr}. \tag{58}$$

The first term in Eq. (58), $\varepsilon_{0Ir}$ is the imaginary relative permittivity of absorption away from the mass $r \to \infty$ or when $M \to 0$, identical to that obtained previously, Eq. (46). The second term, $\varepsilon_{0gIr}$, is the imaginary relative permittivity of absorption induced by the mass $M$ at a distance $r$

$$\varepsilon_{0gIr} = \frac{8\pi GM N_0 e^2 \omega \gamma_{pt}}{m_e c_0^2 r\left[\left(\omega_{0p}^2 - \omega^2\right)^2 + \omega^2\gamma_{pt}^2\right]}. \tag{59}$$

Substituting in Eq. (59), the expression $\gamma_{pt} = \gamma'_p + 2\omega^2 e^2/(3m_e c_0^3)$ and taking into account, according to relation (19), that $4\pi N_0 e^2/(m_e \omega_{0\upsilon}^2) = 1$, we obtain

$$\varepsilon_{0gIr} = \frac{2GM\omega_{0p}^2 \gamma'_{pt} \omega}{c_0^2 r\left[\left(\omega_{0p}^2 - \omega^2\right)^2 + \omega^2\gamma_{pt}^2\right]} + \frac{4GM e^2 \omega_{0p}^2 \omega^3}{3m_e c_0^5 r\left[\left(\omega_{0p}^2 - \omega^2\right)^2 + \omega^2\gamma_{pt}^2\right]} = \varepsilon'_{0gI} + \varepsilon_{0gIsc}. \tag{60}$$

It follows that, for vacuum we distinguish an imaginary permittivity $\varepsilon'_{0gI}$ of absorption of em waves, induced by the gravitational field,

$$\varepsilon'_{0gI} = \frac{2GM\omega_{0p}^2 \gamma'_p \omega}{c_0^2 r \left[\left(\omega_{0p}^2 - \omega^2\right)^2 + \omega^2 \gamma_{pt}^2\right]} \tag{61}$$

and an imaginary permittivity determined by the scattering of emw, $\varepsilon_{0gIsc}$

$$\varepsilon_{0gIsc} = \frac{4GMe^2 \omega_{0p}^2 \omega^3}{3m_e c_0^5 r \left[\left(\omega_{0p}^2 - \omega^2\right)^2 + \omega^2 \gamma_{pt}^2\right]}. \tag{62}$$

These imaginary permittivities correspond to an attenuation coefficient and an attenuation distance [25], according to relations (47):

$$\beta'_g = \frac{2\omega \varepsilon'_{0gI}}{c_0} = \frac{4GM \omega_{0p}^2 \gamma'_p \omega^2}{c_0^3 r \left[\left(\omega_{0p}^2 - \omega^2\right)^2 + \omega^2 \gamma_{pt}^2\right]}, \quad D'_g = \frac{1}{\beta'_g} = \frac{c_0^3 r \left[\left(\omega_{0p}^2 - \omega^2\right)^2 + \omega^2 \gamma_{pt}^2\right]}{4GM \omega_{0p}^2 \gamma'_p \omega^2}; \tag{63}$$

$$\beta_{gsc} = \frac{2\omega \varepsilon_{0gIsc}}{c_0} = \frac{8GMe^2 \omega_{0p}^2 \omega^4}{3m_e c_0^6 r \left[\left(\omega_{0p}^2 - \omega^2\right)^2 + \omega^2 \gamma_{pt}^2\right]}, \quad D_{gsc} = \frac{1}{\beta_{gsc}} = \frac{3m_e c_0^6 r \left[\left(\omega_{0p}^2 - \omega^2\right)^2 + \omega^2 \gamma_{pt}^2\right]}{8GMe^2 \omega_{0p}^2 \omega^4}. \tag{64}$$

Since $\gamma'_p \ll 2\omega^2 e^2 / (3m_e c_0^3)$, it follows the $\gamma_{pt} \cong 2\omega^2 e^2 / (3m_e c_0^3)$ and we can approximate expressions (63, 64) in the form:

$$\beta'_g = \frac{4GM \omega_{0p}^2 \gamma'_p \omega^2}{c_0^3 r \left[\left(\omega_{0p}^2 - \omega^2\right)^2 + 4\omega^6 e^4 / (9m_e^2 c_0^6)\right]}, \quad D'_g = \frac{1}{\beta'_g} = \frac{c_0^3 r \left[\left(\omega_{0p}^2 - \omega^2\right)^2 + 4\omega^6 e^4 / (9m_e^2 c_0^6)\right]}{4GM \omega_{0p}^2 \gamma'_p \omega^2}; \tag{65}$$

$$\beta_{gsc} = \frac{8GMe^2 \omega_{0p}^2 \omega^4}{3m_e c_0^6 r \left[\left(\omega_{0p}^2 - \omega^2\right)^2 + \frac{4\omega^6 e^4}{9m_e^2 c_0^6}\right]}, \quad D_{gsc} = \frac{1}{\beta_{gsc}} = \frac{3m_e c_0^6 r \left[\left(\omega_{0p}^2 - \omega^2\right)^2 + \frac{4\omega^6 e^4}{9m_e^2 c_0^6}\right]}{8GMe^2 \omega_{0p}^2 \omega^4}. \tag{66}$$

Since the attenuation coefficient (by absorption and scattering) in the gravitational field is inversely proportional to the distance, $\beta \sim 1/r$, the attenuation $\beta r$ is a constant (it does not depend on $r$, but is a function of $M$ and $\omega$).

For small angular frequencies, $\omega \ll \omega_{0\upsilon} = 2m_e c_0^2/\hbar$, these quantities become:

$$\beta'_g \cong \frac{4GM \gamma'_p \omega^2}{\omega_{0p}^2 c_0^3 r}, \quad D'_g \cong \frac{m_e^2 c_0^7 r}{GM \hbar^2 \gamma'_p \omega^2}, \quad \lim_{\omega \to 0} D'_g = \infty; \tag{67}$$

$$\beta_{gsc} \cong \frac{8GMe^2 \omega^4}{3m_e c_0^6 \omega_{0p}^2 r}, \quad D_{gsc} \cong \frac{3m_e c_0^6 \omega_{0p}^2 r}{2GMe^2 \omega^4} = \frac{3m_e^3 c_0^{10} r}{8GMe^2 \hbar^2 \omega^4}, \quad \lim_{\omega \to 0} D_{gsc} = \infty. \tag{68}$$

For frequencies close to the angular resonance frequency, $\omega \cong \omega_{0\upsilon} = 2m_e c_0^2/\hbar$, with $r_{gM} = GM/c_0^2$, these quantities become:

$$\beta'_{g0} \cong \frac{9m_e^2 c_0^3 GM \gamma'_p}{\omega_{0p}^2 e^4 r}, \quad D'_{g0} \cong \frac{4e^4 c_0 r}{9GM \hbar^2 \gamma'_p} = \left(\frac{e^2}{\hbar c_0}\right)^2 \frac{4c_0 r}{9r_{gM} \gamma'_p} = D'_0 \frac{r}{r_{gM}}; \tag{69}$$

$$\beta_{g0sc} \cong \frac{6GMm_e}{e^2 r}, \quad D_{g0sc} \cong \frac{e^2 r}{6GMm_e} = \frac{r_e r}{6r_{gM}} = D_{0sc} \frac{r}{r_{gM}}. \tag{70}$$

For high angular frequencies, $\omega \gg \omega_{0\upsilon} = 2m_e c_0^2/\hbar$, these quantities become:

$$\beta'_g \cong \frac{9m_e^2 c_0^3 GM \omega_{0p}^2 \gamma'_p}{e^4 \omega^4 r}, \quad D'_g \cong \left(\frac{e^2}{\hbar c_0}\right)^2 \left(\frac{\hbar\omega}{m_e c_0^2}\right)^4 \frac{c_0 r}{18 r_{gM} \gamma'_{pt}} = D' \frac{r}{r_{gM}}, \quad \lim_{\omega \to \infty} D'_g = \infty; \quad (71)$$

$$\beta_{gsc} = \frac{6GM m_e^3 c_0^4}{e^2 \hbar^2 \omega^2 r}, \quad D_{gsc} = \frac{e^2 \hbar^2 \omega^2 r}{6GM m_e^3 c_0^4} = \left(\frac{\hbar\omega}{m_e c_0}\right)^2 \frac{r_e r}{12 r_{gM}} = D_{sc} \frac{r}{r_{gM}}, \quad \lim_{\omega \to \infty} D_{gsc} = \infty. \quad (72)$$

Verification of these frequency dependencies will be possible in the future, through the improvement of measuring devices.

## 5. Dependence of the external relative permittivity of the vacuum on the angular frequency

According to the expression (9) of the real relative permittivity, written for vacuum, for $\omega > 0$ this it also depends on the angular frequency

$$\varepsilon_{Rr} = \frac{N_0 q_e^2 \left(\omega_{0\upsilon}^2 - \omega^2\right)}{\varepsilon_0 m_e \left[\left(\omega_{0\upsilon}^2 - \omega^2\right)^2 + \omega^2 \gamma_t^2\right]} = \frac{\pi N_0 e^2 \hbar^2}{m_e^3 c_0^4 \left(1 - \omega^2/\omega_{0\upsilon}^2\right)\left[1 + \omega^2 \gamma_t^2/\omega_{0\upsilon}^4 \left(1 - \omega^2/\omega_{0\upsilon}^2\right)\right]}. \quad (73)$$

Substituting Eq. (22) and the expression $\gamma_{pt} \cong \gamma_p \cong 2\omega^2 e^2 / (3m_e c_0^3)$ in Eq. (73), we get

$$\varepsilon_{Rr} = \frac{N_0 q_e^2 \left(\omega_{0\upsilon}^2 - \omega^2\right)}{2\varepsilon_0 m_e \left[\left(\omega_{0\upsilon}^2 - \omega^2\right)^2 + \omega^2 \gamma_t^2\right]} = \frac{\pi N_0 e^2 \hbar^2}{2m_e^3 c_0^4 \left(1 - \omega^2/\omega_{0\upsilon}^2\right)\left[1 + \omega^2 \gamma_t^2/\omega_{0\upsilon}^4 \left(1 - \omega^2/\omega_{0\upsilon}^2\right)\right]} \cong \\ \frac{1}{\left(1 - \omega^2/\omega_{0\upsilon}^2\right)\left[1 + 16\alpha^2 \omega^6/9\omega_{0\upsilon}^6 \left(1 - \omega^2/\omega_{0\upsilon}^2\right)^2\right]}. \quad (74)$$

At low frequencies, $\omega \ll \omega_{0\upsilon} = 2m_e c_0^2/\hbar$, the external relative permittivity of the vacuum (74) is approximated by

$$\varepsilon_{Rr} \cong 1 + \frac{\omega^2}{\omega_{0\upsilon}^2}. \quad (75)$$

For the visible spectrum, $\omega \cong 3 \cdot 10^{15} \mathrm{s}^{-1}$, the relative permittivity is $\varepsilon_{Rr} = 1 + (\omega/\omega_{0\upsilon})^2 \cong 1 + 10^{-12}$. By increasing the frequency, the relative permittivity increases and reaches a positive maximum for the frequency $\omega_{0+} \cong (1 - 2\alpha/\sqrt{3})\omega_{0\upsilon}$. For frequencies higher than this the permittivity decreases and at the frequency $\omega_1 \cong (1 - 8\alpha^2/9)\omega_{0\upsilon}$ becomes equal to unity. At the frequency equal to the angular resonance frequency, $\omega = \omega_{0\upsilon} = 2m_e c_0^2/\hbar$, the relative permittivity becomes zero. For higher frequencies, $\omega \gg \omega_{0\upsilon} = 2m_e c_0^2/\hbar$, the permittivity becomes negative and reaches an extremum at the frequency $\omega_{0-} \cong (1 + 2\alpha/\sqrt{3})\omega_{0\upsilon}$ and then, for higher frequencies, decreases and reaches the zero limit when $\omega \to \infty$. This type of frequency dependence is only partially confirmed, in relation to the frequency dependence of the gravitational index [56, 57], by experimental studies and requires more intensive experimental research. The discrepancy with the experimental results indicates that the expression of the frequency dependence of the vacuum permittivity is partially correct and requires corrections. These corrections can be obtained by studying the frequency dependence of the permittivity of the permeability by taking into account the mutual influence of the two intense and rapidly oscillating electric and magnetic fields [58] on the motion of the fermion pairs.

## 6. Relative permittivity of vacuum in electrostatic field

According to quantum physics, the relative permittivity of a vacuum under the action of an electromagnetic field is given by the relation deduced by Euler and Kockel [41, 24 - Linear Superposition]. For an electrostatic field with energy density $w_E = \varepsilon_0 E^2/2$, this relation is

$$\varepsilon_{0rE} = 1 + \frac{2^4 e^4 \hbar w_E}{45 m_e^4 c_0^7} = 1 + \frac{2^4 \alpha^2 \hbar^3 w_E}{45 m_e^4 c_0^5}. \tag{76}$$

We can deduce this relationship by applying the method used to obtain the expression of the relative gravitational permittivity, as an effect of the change in the number of e-p pairs under the action of the electrostatic field. The electrostatic field polarizes the vacuum and acts on the e-p dipoles which have potential energy $W_E = -\vec{p}_{e-p}\vec{E} = \left[-q_e^2/(m_e\omega_0^2)\right]E^2$ [24- Ch. 4]. The Boltzman distribution of the pair density in the vacuum becomes

$$N_{E\upsilon} \cong N_0 \exp\left(\frac{-W_{e\upsilon}}{kT_{0\upsilon}}\right) \cong N_0\left(1 + \frac{q_e^2 E_\upsilon^2}{m_{e\upsilon}\omega_{0\upsilon}^2 kT_{0\upsilon}}\right). \tag{77}$$

If we consider that the e-p dipoles are in the inner vacuum with relative permittivity $\varepsilon_{r\upsilon}$ and applying the dependence given by the W-D theory of the physical parameters on this index, we obtain: $m_{e\upsilon} = m\varepsilon_{r\upsilon}^{3/2}$, $kT_{0\upsilon} = kT_0\varepsilon_{r\upsilon}^{-1/2} = m_e c_0^2 \varepsilon_{r\upsilon}^{-1/2}$, $E_\upsilon = E$ and $\omega_{0\upsilon}^2 = 4m_e^2 c_0^4/\hbar^2$. Replacing these dependencies in expression (77), with those given by expression (15), we obtain

$$N_{E\upsilon} \cong N_0\left(1 + \frac{2e^2 \hbar^2 w_E}{m_e^4 c_0^6 \varepsilon_{r\upsilon}}\right) = N_0\left(1 + \frac{2^7 \hbar^3 w_E}{m_e^4 c_0^5 \alpha^3}\right). \tag{78}$$

With this expression for the pair density, the relative permittivity of vacuum (22) modified by interaction with the electrostatic field is

$$\varepsilon_{0rE} = \frac{\pi N_{E\upsilon} e^2 \hbar^2}{m_e^3 c_0^4} = \frac{\pi N_0 e^2 \hbar^2}{m_e^3 c_0^4}\left(1 + \frac{2^7 \hbar^3 w_E}{m_e^4 c_0^5 \alpha^3}\right) = 1 + \frac{2^7 \hbar^3 w_E}{m_e^4 c_0^5 \alpha^3}. \tag{79}$$

This expression differs from that of the relative permittivity deduced in quantum physics (76), $2^7 \hbar^3 w_E/(m_e^4 c_0^5 \alpha^3) = \left[2^4 \alpha^2 \hbar^3 w_E/(45 m_e^4 c_0^5)\right](2^3 45/\alpha^5) \gg 2^4 \alpha^2 \hbar^3 w_E/(45 m_e^4 c_0^5)$. We can correct this expression, if we introduce the averaged dipole energy taking into account the decay probability used by Mainland in his papers, $\langle W_E \rangle = W_E(N_0 \Gamma_{ep}\Delta_{ep}/N_0) = W_E(\alpha^5/4)$. With this correction the relative permittivity becomes $\varepsilon_{0rE} - 1 = 2^5 \hbar^3 \alpha^2 w_E/(m_e^4 c_0^5) > 2^4 \alpha^2 \hbar^3 w_E/(45 m_e^4 c_0^5)$. The discrepancy between the two expressions of the relative permittivity of the vacuum induced by the electrostatic field signals that the W-D hypothesis used in deducing the expressions for the energy of the dipoles in the vacuum, $W_E$, in the external electrostatic field and the thermal energy, $kT_{0\upsilon}$, of the dipoles is not correct. It is possible that the internal vacuum with relative permittivity $\varepsilon_{r\upsilon}$ behaves like a regular dielectric (it affects only the propagation speed of the interactions and therefore the intensity of the fields and not the etalons of length, time, mass-energy, etc.). The problem of physical phenomena in a dielectric (whether the dielectric affects the gravitational interaction and the etalons) has been studied [59, 60] in connection with nonlinear phenomena at high intensities of the electromagnetic field. These studies, which are a development and application of the results of the Euler-Kockel papers, showed that the intense electromagnetic field modifies the optical properties of the vacuum, analogous to a gravitational field, and can introduce a geometrization of space-time around a body with mass and electric charge.

According to GRT, the metric near an electron (without rotation) is the Reissner-Nordstrom metric [61, 62] which is equivalent to a relative permittivity $\varepsilon_{0ge} = 1 + 2r_{ge}/r - 2r_{ge}r_e/r^2$, $r_{ge} = Gm_e/c^2$, $r_e = e^2/(m_e c^2)$ [63]. For the electron with spin, the metric is of the Kerr–Newman type. This metric includes in the expression of the relative permittivity also a length (radius) determined by the electron's angular momentum $a = \hbar/(2m_e c_0)$ [64]. If we complete the expression of this relative permittivity with the component determined by the nonlinearity of the electron's electromagnetic field, according to the papers [59, 60], $\left[2^3 e^4 \hbar/(45 m_e^4 c_0^7)\right](2w_E + 5w_B) = \left[2^3/(45\alpha)\right]\left(r_e^4/e^2\right)\left[e^2/(4\pi r^4) + 5w_B\right]$, $w_B = B^2/(2\mu_0)$ (including the energy density of the electron's magnetic field is more difficult because, at distances smaller than the Compton length, for the point model, we do not know how it depends on the electron's magnetic moment and radius) we obtain the permittivity corresponding to a metric that would completely describe an extended particle [65-67] with charge and spin. The need for a particle model with electric charge and spin is also found in the model of a particle in the Acoustic World [68].

## 7. Comments, discussions and conclusions

Resuming the analysis of dielectric and magnetic properties of the physical vacuum was determined by the papers published in recent years and which confirmed the hypothesis of the vacuum as a material system with physical properties. The importance of this hypothesis is greater when combined with the possibility of physically modeling the inference of the gravitational refractive index induced by a mass. The phenomenon of the change in the refractive index of the physical vacuum by a gravitational field is not surprising if we take into account the fact that in quantum physics the change in the refractive index is inferred from the interaction with an electromagnetic field [41-43, 24].

The results obtained in the cited papers [1, 3, 4, 11, 12, 14-17, 27, 42-43], similar in initial hypothesis (polarization of vacuum modeled as a material system) and different in solution, are approximate because a semiclassical theory is used. For this reason, a rigorous quantum approach is necessary to obtain better results. We note that the approximate solution of the expression for the vacuum index does not affect the derivation of the expression for the gravitational refractive index which depends on the Boltzman distribution. It is possible that the deviation from unity of the gravitational index is only half of the value accredited by general relativity, the other half coming from the interaction of photons as particles with motion mass ($m_\omega = \hbar\omega/c^2$) [69, 70]. The rigorous quantum approach is also necessary to explain the other two properties of the physical vacuum modeled as a material system: superconductivity (the existence of displacement current generated by variable electric fields in capacitors placed in vacuum and the absence of thermal effects) and superfluidity (to explain the absence of the viscosity phenomenon). These phenomena involve the formation of Cooper pairs [71] from fermion-antifermion pairs and the behavior of this fluid of Cooper pairs as a Bose-Einstein condensate [72].

Modeling a non-point particle with charge, mass and spin can follow both the quantum and general relativistic path, but also the phenomenological path of inducing the vacuum index in the presence of such a particle. The results are not mutually exclusive but complementary.